\documentclass[twocolumn,amsmath,amssymb,pra]{revtex4-2}

\usepackage{txfonts}
\usepackage{graphicx}
\usepackage{dcolumn}
\usepackage{bm}
\usepackage{color}
\usepackage{braket}

\usepackage[colorlinks=true,linkcolor=blue,urlcolor=blue,
anchorcolor=blue,citecolor=blue,bookmarksnumbered]{hyperref}

\usepackage{amsmath,amssymb}
\usepackage{graphicx}
\begin{document}
\title{Exploring vibronic dynamics near a sloped conical intersection with trapped Rydberg ions}	
\author{Abdessamad Belfakir$^1$}
\author{Weibin Li$^2$}
\affiliation{$^1$ The UM6P Vanguard Center, Mohammed VI Polytechnic University (UM6P), 
Rocade Rabat-Salé, Technopolis, 11103 Morocco\\
$^2$ School of Physics and Astronomy, and Centre for the Mathematics 
and Theoretical Physics of Quantum Non-equilibrium Systems,
University of Nottingham, Nottingham, NG7 2RD, UK}
\begin{abstract}
 We study spin-phonon coupled dynamics in the vicinity of a sloped conical intersection created by laser coupling the electronic (spin) and vibrational degrees of freedom of a pair of trapped Rydberg ions. We show that the shape of the potential energy surfaces can be engineered and controlled by exploiting the sideband transitions of the crystal vibration and dipole-dipole interactions between Rydberg ions in the Lamb-Dicke regime.  Using the sideband transition, we realize  a sloped  conical intersection whose cone axis is only tilted along one spatial axis. When the phonon wavepacket is located in the potential minimum of the lower potential surface, the spin and phonon dynamics are largely frozen owing to the geometric phase effect. When starting from the upper potential surface, the electronic and phonon states tunnel to the lower potential surface, leading to a partial revival of the initial state. In contrast, the dynamics drastically change when the initial wavepackets are away from the conical intersection. The initial state is revived, and is almost entirely irrelevant to whether  it is from the lower or upper potential surface. Complete Rabi oscillations of the adiabatic states are found when the wavepacket is initialized on the upper potential surface. The dynamics occur on the microsecond and nanometer scales, implying that Rydberg ions provide a platform for simulating nonadiabatic processes in the vicinity of a sloped conical intersection.
	\end{abstract}
\maketitle
\section{Introduction }\label{sec1}
A conical intersection (CI) is a point in the nuclear coordinate space where two or more Born-Oppenheimer potential-energy surfaces (PESs) are degenerate \cite{Yarkony2,Baer,Domcke}.  CIs are characterized by a linear dependence on the nuclear coordinates and associated geometric effects, which are absent in other forms of degeneracy, such as glancing intersections, which depend  quadratically on the coordinates and do not exhibit geometric effects~\cite{Mozhayskiy,Lee}. The electronic and nuclear wavefunctions accumulate a  phase factor of $\pi$ during their motion around a path that encircles a CI \cite{Bohm,Longuet-Higgins,Berry,Mead,Althorpe,Ryabinkin,Ryabinkin2}. CIs appear in
many photochemical processes in
molecules \cite{Yarkony,Worth}. Due to this degeneracy, they enable ultrafast and radiationless decay of intramolecular excited states which grants, for example, the stability of DNA \cite{Barbatti}, vision mechanism \cite{Schoenlein,Polli,Rinaldi} and  photosynthesis processes \cite{Hammarstroem}.  CIs and geometric (Berry) phases are also important in solid-state systems~\cite{larson2020conical}. 
\begin{figure}
\centering
\includegraphics[width=0.9\columnwidth]{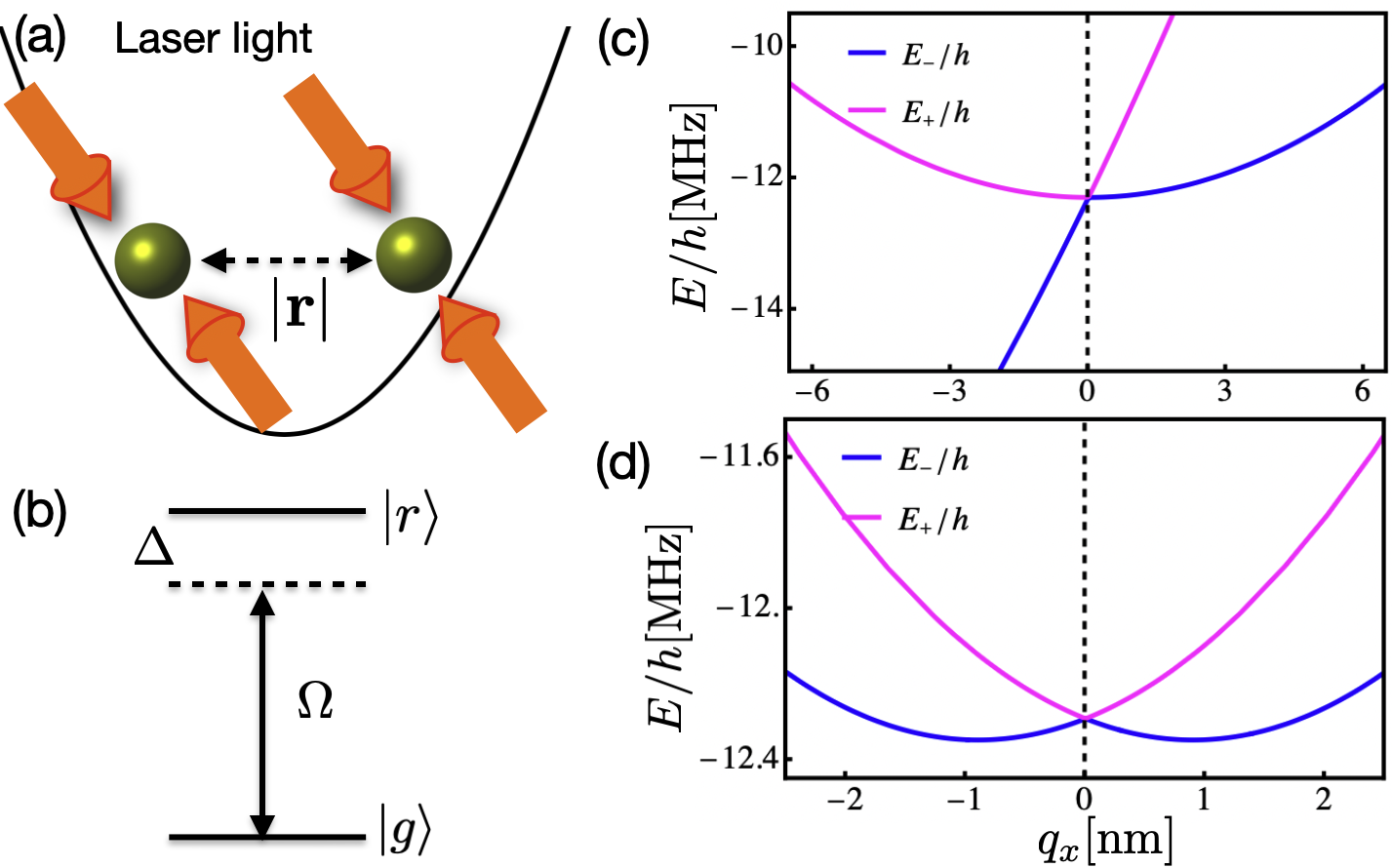}
\caption{(a) CI in a system of two trapped Rydberg ions using the sideband resolved laser excitation. (b) Each Rydberg ion is modeled
as a two-level system, laser excited from the ground state $\ket{g}$ to the Rydberg state $\ket{r}$. The two ions are coupled to a standing wave laser $E(\Omega_{0},\Delta)$ with  Rabi frequency $\Omega_{0}$ and detuning $\Delta$. The dipole-dipole interaction  $V_{DD}$ couples the vibration to the Rydberg state. (c) Sloped CIs. (d) Peaked CI, where the upper PES is entirely above the lower PES in the vicinity of the CI.   In (c) and (d), the profiles of the CI  along $Q_y=0$ are shown. Here we consider Sr$^+$ ions with mass $m=87.9\times1.66\times 10^{-27}\,\text{Kg}$, $\omega_x=2\pi\,\text{MHz}$, $\omega_y=2\pi\times 1.6  \,\text{MHz}$, $F_0/\hbar=- 1441.99\,\text{MHz/{$\mu$m}}$, $G_0/\hbar=88.8572\,\text{MHz/{$\mu$m}}$ and  $\Delta/\hbar=-12.29\,\text{MHz}$.  The position of the CI is $(Q_y^*,q_x^*)=(0,0)$. }
	\label{fig_diagram}
\end{figure}

Molecular dynamics near CIs
typically occur on femtosecond timescales, and thermal effects are vital during these dynamics \cite{Ismail,Perun}. Detecting and observing features associated with a CI therefore require
temporal and broad spectral resolutions, as well as the control of the nuclear wavefunction  \cite{Nunn,Schoenlein, Kowalewski,Adachi,Young}. Numerical simulations of the CI dynamics of  real molecules are challenging because of the extremely high dimensions of the associated Hilbert spaces~\cite{Tuckerman}. The Born-Oppenheimer (BO) approximation, where  the nuclei are supposed to move only on a single
PES, rather than moving on all PESs, fails to study the dynamics of the nuclei near a CI, for example, the geometric effects~\cite{Tully}.

{ In recent years, highly controllable quantum systems have been developed to investigate CI dynamics. Direct observation of CIs has been realized with single trapped ions \cite{Valahu2023,Whitlow2023}, and wavepacket branching near a CI has been reported in circuit QED \cite{PhysRevX.13.011008}. These experiments created CIs by coupling vibrational (phonon) and electronic (spin) states in the sideband-resolved regime. In a recent work \cite{Gambetta}, it has been shown that a CI can be created between a pair of trapped Rydberg ions.  Rydberg ions have strong dipolar exchange interactions \cite{Li2014,PhysRevLett.125.133602,Zhang2020}, high Rydberg states polarizabilities \cite{Schmidt-Kaler2011,PhysRevLett.108.023003}, and long lifetimes \cite{Muller2008,Mokhberi_2020}. Owing to these properties, ion crystal vibrations and electronic states strongly interact in a Paul trap, as demonstrated in a recent experiment~\cite{mallwegerProbingElectronicStatedependent2025}. To create a CI, the minimal requirement is that two phonon modes are coupled to two different spin operators. In Ref.~\cite{Gambetta} a pair of Rydberg ions where one is in a Rydberg $nP$ state and the other in $nS$ state ($n$ is the principal quantum number) were prepared.  The two Rydberg states couple the crystal vibration along the trap axis through spatial gradients of dipolar exchange interactions. To induce the second coupling between a different spin operator and vibration along a different direction, an electric field is applied to shift the ion crystal off the trap axis. Using the state-dependent trap potential \cite{PhysRevA.87.052304}, this creates a state-dependent coupling with phonons perpendicular to the trap axis. Using the engineered symmetric (peaked) CI, dynamics and geometric effects are explored in the nanometer and microsecond scales \cite{Gambetta}.} \\

In this work, we investigate the diabatic spin-phonon coupled dynamics around a \textit{sloped} CI with the trapped Rydberg ion quantum simulator. Our setting is a pair of  Rydberg ions trapped in a linear Paul trap, as depicted in Fig.~\ref{fig_diagram}(a). The main difference between the present scheme and the previous one~\cite{Gambetta} is that, here, the electronically low-lying 
state $\ket{g}$ and the Rydberg state $\ket{r}$ constitute the spin 
degrees of freedom, with ions in the Rydberg state $\ket{r}$ interacting 
strongly via a density--density interaction~\cite{Li2014}.
 One type of spin-phonon coupling is induced through the slope of the Rydberg ion interaction along the trap $x$-axis, leading to two-ion spin-phonon coupling. A laser induced spin-dependent force \cite{PhysRevLett.92.207901,PhysRevA.72.063407}
 is applied along the $y$-axis. Under the anti-blockade condition and in the Lamb-Dicke regime~\cite{Leibfried_2003, Cai_2021_NJP,Monroe_2021}, this leads to the required two-ion spin-phonon coupling perpendicular to the previous one.  Since the static electric field~\cite{Gambetta} is not used in this scheme, we avoid the additional spin–phonon coupling that would otherwise be induced by the field. In this setting, we create a CI whose slopes differ along {one of the two axes} [Fig.~\ref{fig_diagram}(c)]. In contrast, in a peaked (symmetric) CI, the slopes are identical in all directions [Fig. ~\ref{fig_diagram}(d)]. We then investigate the diabatic and adiabatic dynamics and the nuclear densities in the neighborhood of the constructed sloped CI. When the system is initially prepared at the minimum of the lower PES, we observe freezing of the system wavepacket. This can be explained by the geometric phase effect. However, when the phonon wavepacket is initially prepared on the upper PES, the phonon and electronic states tunnel to the other PES, leading to a partial revival of the initial wavepacket. A complete revival of the initial wavepacket is observed when the system is prepared away from the CI, and this does not depend on whether we prepare the system on the lower PES or the upper PES. 
 
 {This paper is organized as follows. In Sec.~\ref{The model and equations of motion}, we present the physical setup and introduce the model that generates a sloped CI using two trapped Rydberg ions. We also derive the effective Hamiltonian, discuss the resulting PESs, and characterize the topography of the CI. In Sec.~\ref{Nuclear motion in  neighbourhood of a CI}, we study the dynamics of the diabatic and adiabatic states, as well as the evolution of the nuclear densities, for different initial conditions. Finally, Sec.~\ref{Conclusions} summarizes our main findings and outlines possible future directions.}
\section{Model of two trapped ions}\label{The model and equations of motion}
\subsection{The system and Hamiltonian}
We consider two Rydberg ions of mass $m$  in a linear
Paul trap that applies an effective time-independent harmonic potential with trapping frequencies $\omega_\xi  $ $(\xi=x,y,z)$. In the following we assume that $\omega_z\gg \omega_x,\omega_y$, ensuring that the two ions are confined in the $x$-$y$ plane. The trapping frequencies are determined by the electric-field gradients of both the oscillating and static quadrupole fields, as well as by the radio-frequency of the Paul trap~\cite{Leibfried_2003,Higgins,Single,Major}. The effective trapping potential of the two ions is given by,
\begin{equation}
	V_{\text{trap}}=\dfrac{m}{2}\big[\omega_x^2 (X_1^2+X_2^2)+\omega_y^2 (Y_1^2+Y_2^2)\big]+\mathcal{K}e^2/r,
\end{equation}
where $\textbf{R}_1=(X_1,Y_1)$ and  $\textbf{R}_2=(X_2,Y_2)$ stand for the nuclear coordinates  of the two ions. Coefficient $\mathcal{K}$ is the Coulomb's constant, $e$ is the elementary charge and $r=|\textbf{r}|=|\textbf{R}_2-\textbf{R}_1|$  {is the inter-ion separation.}  For concreteness, we consider $\mathrm{Sr}^+$ ions, although the scheme is equally applicable to $\mathrm{Ca}^+$ ions with appropriate choices of the laser and trap parameters. Experimental realizations of Rydberg excitation in both $\mathrm{Ca}^+$~\cite{PhysRevLett.115.173001,PhysRevLett.127.203001} and $\mathrm{Sr}^+$ ions ~\cite{PhysRevX.7.021038,PhysRevLett.119.220501} have been demonstrated.

The two ions are confined along the $x$-axis. Each ion is modeled as a two-level system, driven by a laser that couples the ground state $\ket{g_j}$ to the Rydberg state $\ket{r_j}$ with Rabi frequency $\Omega_0$ and detuning $\Delta$ ($j = 1,2$). 
The excitation laser forms a standing wave with wave number $k$ along the $y$-axis. The ions are located at the nodes of the standing wave laser.
In the rotating frame and using $\hbar \equiv 1$, the interaction Hamiltonian is,
\begin{equation}
    \label{Hamiltonian_sum}
    H_R = \Delta(\hat{n}_1 + \hat{n}_2)
    + \frac{\Omega_0}{2}\big[ \sin(k Y_1)\,\sigma_1^x 
    + \sin(k Y_2)\,\sigma_2^x \big],
\end{equation}
where $\sigma_j^x = \ket{g_j}\bra{r_j} + \ket{r_j}\bra{g_j}$ and 
$\hat{n}_j = \ket{r_j}\bra{r_j}$~\cite{Cai_2021_NJP,Chung_PRA_2023}.

We  work in the  Lamb–Dicke regime at low temperatures. Hence, by expanding  $\sin{(kY_j)}$ to the first order of the corresponding Lamb-Dicke  parameters, the ion-laser interaction becomes,
\begin{equation}
	\label{Hamiltonian_sum}
	H_R\approx \Delta( \hat{n}_1+\hat{n}_2)+G_0( \delta Y_1\,\sigma_1^x+ \delta Y_2\, \sigma_2^x),
\end{equation}
where $G_0\equiv\dfrac{{k}\Omega_{0}}{2}$ \cite{Cai_2021_NJP,Chung_PRA_2023} characterizes the coupling strength between an ion and the laser, and $\delta Y_j = Y_j-Y_j^0$ denotes the deviation from the equilibrium position $Y_j^0$. The dipole-dipole interaction in the Rydberg state is,
\begin{equation}
	H_{\text{DD}}=V_{\text{DD}}\times \hat{n}_1\otimes\hat{n}_2,
\end{equation}
where
$
	V_{\text{DD}}=C_0/|X_2-X_1|^3,
$
with $C_0$ being the dispersion constant. The full Hamiltonian is given by,
\begin{equation}\label{Hamiltonian_full}
	H=\big(T+V_{\text{trap}}\big)+ H_R+H_{\text{DD}},
\end{equation}
where $T=-\dfrac{\nabla_{{R_1}}^2}{2m}-\dfrac{\nabla_{{R_2}}^2}{2m}$ is the kinetic operator.
We analyze the dynamics in both the center-of-mass (CM) and  relative frames, where the CM coordinate is  $\mathbf{R} =(X,Y)= \frac{\mathbf{R}_1 + \mathbf{R}_2}{2}$
and the relative coordinate is
    $\mathbf{r}=(x,y) = \mathbf{R}_2 - \mathbf{R}_1$.
In these coordinates, the kinetic energy operator becomes,
\begin{equation}
    T = -\frac{\nabla_{\mathbf{R}}^{2}}{2M} - \frac{\nabla_{\mathbf{r}}^{2}}{2\mu},
\end{equation}
and the dipole--dipole interaction is,
\begin{equation}
    H_{\text{DD}} = \frac{C_0}{|x|^{3}}\, \hat{n}_1 \otimes \hat{n}_2,
\end{equation}
where $M = 2m$ and $\mu = m/2$.
In terms of the CM and relative coordinates, the trapping potential can be written as,
\begin{eqnarray}
    V_{\text{trap}} &=& 
    \frac{M}{2}\big(\omega_x^2 X^2 + \omega_y^2 Y^2\big)
    + \frac{\mu}{2}\big(\omega_x^2 x^2 + \omega_y^2 y^2\big)
    + \frac{\mathcal{K} e^2}{\sqrt{x^2 + y^2}}. \nonumber
\end{eqnarray}

{From the potential energy $V_{\text{trap}}$,  the equilibrium positions $\textbf{R}_0=(X_0,Y_0)=(0,0)$ and $\textbf{r}_0=(x_0,0)$ are obtained 
 by numerically solving $
\left. \frac{\partial V_{\text{trap}}}{\partial\textbf{R}} \right|_{\textbf{R}=\textbf{R}_0} = 0
$ and $
\left. \frac{\partial V_{\text{trap}}}{\partial \textbf{r}} \right|_{\textbf{r}=\textbf{r}_0} = 0, $ respectively
\cite{James1998}. }

As the ions oscillate around their equilibrium separation, we expand the dipole--dipole interaction $H_{\text{DD}}$ about the equilibrium distance $x_0$, yielding
\begin{equation}\label{Vex}
    H_{\text{DD}} = V_0\, \hat{n}_1 \otimes \hat{n}_2
    - F_0 q_x\, \hat{n}_1 \otimes \hat{n}_2,
\end{equation}
where $V_0 = \dfrac{C_0}{x_0^{3}}$ denotes the interaction strength at equilibrium.
Spatial deviations from equilibrium, $q_x = x - x_0$, couple the Rydberg excitation to the vibrational motion through the slope of the interaction potential,
\[
    F_0 = \frac{3C_0}{x_0^{4}}.
\]
Spin-phonon coupling is an essential ingredient for generating the CI.
  
In the following, we obtain the second spin--phonon coupling through the laser--ion interaction, which acts along different spin and spatial directions. 
We choose the detuning such that $\Delta + V_0 = 0$, and  consider a weak laser field satisfying 
$|G_0 \delta Y_1|, |G_0 \delta Y_2| \ll |\Delta|$. 
Under these conditions, the pair states 
$\{\ket{rr},\, \ket{+} = (\ket{rg} + \ket{gr})/\sqrt{2}\}$ 
are resonantly coupled by the laser field, whereas the pair state $\ket{gg}$ is far detuned and can be adiabatically eliminated from the dynamics of the system. 
The antisymmetric state $\ket{-} = (\ket{rg} - \ket{gr})/\sqrt{2}$ is decoupled due to parity symmetry. 
We therefore restrict the dynamics to the pair-state subspace $\{\ket{rr},\, \ket{+}\}$, 
which can be prepared, for example, by initializing the ions in the state $\ket{+}$.
The pair $\{\ket{rr},\, \ket{+}\}$ forms a diabatic basis. Combining the laser--ion coupling with the dipole--dipole interaction, we obtain the Hamiltonian,
\begin{eqnarray}
\label{He_Hd}
    H_R &+& H_{\text{DD}} = 
    \Delta\big( \ket{+}\bra{+} + \ket{rr}\bra{rr} \big) \\
    &+& \,{\sqrt{2}}\,G_0 (\delta Y_1+\delta Y_2) \big( \ket{+}\bra{rr} + \ket{rr}\bra{+} \big)
    - F_0 q_x\,\ket{rr}\bra{rr}. \nonumber
\end{eqnarray}

The second line of Eq.~(\ref{He_Hd}) shows that two different types of spin-phonon coupling are obtained. In the first term,   the \textit{CM mode} along the $Y$-axis couples the pair flipping $\ket{+}\to \ket{rr}$ and vice-versa. In the second term, the \textit{breathing mode} (relative motion) along the $x$ axis couples to the pair Rydberg state $\ket{rr}$~\cite{James1998}. These two couplings are used to create a sloped CI.

Introducing the following Pauli matrices,
\begin{eqnarray}
	\label{S_0}
S_0&=&\ket{rr}\bra{rr}+\ket{+}\bra{+},
	\\ \label{Sx}
S_x&=&\ket{+}\bra{rr}+\ket{rr}\bra{+},
	\\ \label{Sz}
		S_z&=&\ket{rr}\bra{rr}-\ket{+}\bra{+},
\end{eqnarray}
which follows that $\ket{rr}\bra{rr}=\dfrac{S_0+S_z}{2}.$

Taking into account only the CM mode along the $y$-axis and the breathing mode along the $x$-axis,  we obtain the desired form of the Hamiltonian
$H = H_0 + H_{\text{ele}}$ with,
\begin{eqnarray}\label{eq:H_0}
    H_0 &=& \left(
        -\dfrac{\nabla_{q_x}^{2}}{2\mu}
        -\dfrac{\nabla_{Q_y}^{2}}{2M}
    \right)\otimes S_0 ,
\end{eqnarray}
and
\begin{equation}\label{eq:H_e}
    H_{\text{ele}} 
    = S(Q,q)\otimes S_0
      + W(Q)\otimes S_x
      + G(q)\otimes S_z .
\end{equation}
The scalar term $S(Q,q)$ has the form,
\begin{equation}\label{S_Q_q}
    S(Q,q)
    = \Delta -\dfrac{F_0\, q_x}{2}
      + \frac{\mu}{2}\bar{\omega}_x^{2} q_x^{2}
      + \frac{M}{2}\omega_y^{2} Q_y^{2}, 
\end{equation}
with $\bar{\omega}_x^{2} = \omega_x^{2} + \dfrac{2\mathcal{K}e^{2}}{\mu |x_0|^{3}}$.
We have also defined $W(Q)= \sqrt{2}\,G_0\, Q_y$ with $Q_y = (\delta Y_1+\delta Y_2)/\sqrt{2}$ and $G(q)= -\dfrac{F_0\, q_x}{2}$.
The Hamiltonian $H$ is one of the important results of this work. We will use $H$ to study the properties and dynamics of the CI. In Ref.~\cite{Gambetta}, the Hamiltonian was engineered through the interplay between the polarizabilities and their strong dipolar exchange interactions, with an external electric field applied to facilitate the coupling.
In the present work, we identify a minimal model that contains a CI using only the dipole--dipole interaction and sideband transitions. The potential advantage is that the Rydberg ions are not perturbed by the electric field, and unwanted spin-phonon coupling can be avoided.
\subsection{The sloped CI}
In the spirit of the Born--Huang approach~\cite{BornHuang1954}, 
the nuclear motion of the ions is governed by the PESs, 
which are obtained as the eigenvalues of the electronic Hamiltonian $H_{\text{ele}}$,  
\begin{equation}\label{PES}
	E_{\pm}=S(Q,q)\pm\sqrt{G(q)^2+W(Q)^2}.
\end{equation}
The corresponding eigenvectors are,
\begin{equation}\label{phi_plus}
\ket{\varphi_{+}(Q,q)}=\cos\big(\Lambda(Q,q)\big)\ket{rr}+\sin\big(\Lambda(Q,q)\big)\ket{+},
\end{equation}
and
\begin{equation}\label{phi_moins}
	\ket{\varphi_{-}(Q,q)}=-\sin\big(\Lambda(Q,q)\big)\ket{rr}+\cos\big(\Lambda(Q,q)\big)\ket{+},
\end{equation}
where $\Lambda(Q,q)$ is determined by $\tan(2\Lambda(Q,q))=W(Q)/G(q)$. These states form the adiabatic basis of the system~\cite{Gambetta}. In Fig.~\ref{fig_diagram}(c), we show the upper and lower PESs, 
$E_{+}$ and $E_{-}$, respectively, for the $^{88}\mathrm{Sr}^{+}$ Rydberg ions. 
A CI occurs when $E_+=E_-$, where the two potential energies cross directly. In the following section, we investigate the nuclear and electronic dynamics and study their influence on the presence of the CI.
 \subsection{Topological properties of the CI}
Near a CI, the topology of the PESs is governed by their linear behavior in the branching plane, spanned by 
 the gradient difference and derivative coupling 
directions~\cite{Yarkony,Lee_2007}. The structure of the CI—whether it is 
symmetric (peaked) or asymmetric and tilted (sloped)—is determined by the 
magnitudes and relative orientation of these directions, together with their 
projections onto seam coordinates. 
The type of the intersection of the PESs depends on the derivatives of the functions $G(q)$ and $W(Q)$. If
\begin{equation}
    \textbf{g}(q_x^*,Q_y^*)=\textbf{$\nabla$}{G(q)}|_{q_x^*,Q_y^*}\neq 0,
\end{equation}
and 
\begin{equation}
    \textbf{h}(q_x^*,Q_y^*)=\textbf{$\nabla$}{W(Q)}|_{q_x^*,Q_y^*}\neq 0,
\end{equation}
the intersection  is conical \cite{Yarkony,Lee_2007}, {where $(q_x^*,Q_y^*)$ denotes the position of the CI}. If $\mathbf{g} = \mathbf{h} = 0$ while second derivatives are nonzero, the 
intersection is of higher order (non-conical) and does not form a CI in the 
usual linear vibronic coupling sense. In order to characterize    the neighborhood of the CI, let us define the following unit vectors,
\begin{equation}
    \textbf{x}=\dfrac{\textbf{g}(q_x^*,Q_y^*)}{g},
\end{equation}
and 
\begin{equation}
    \textbf{y}=\dfrac{\textbf{h}(q_x^*,Q_y^*)}{h},
\end{equation}
where $g=||\textbf{g}(q_x^*,Q_y^*)||$ and $h=||\textbf{h}(q_x^*,Q_y^*)||$. The vectors $\textbf{x}$ and $\textbf{y}$ are the normalized vectors of   the branching plane
vectors $\textbf{g}$ and $\textbf{h}$, respectively \cite{Yarkony,Lee_2007}.  It is also convenient to define,
\begin{equation}
    s_x=\textbf{$\nabla$}{S(Q,q)}|_{q_x^*,Q_y^*}. \textbf{x},
\end{equation}
and 
\begin{equation}
    s_y=\textbf{$\nabla$}{S(Q,q)}|_{q_x^*,Q_y^*}. \textbf{y},
\end{equation}
which are nothing but the projection of the seam coordinates $\textbf{$\nabla$}{S(Q,q)}|_{q_x^*,Q_y^*}$ onto the branching plane
vectors.
The classification of CIs as “peaked” or
“sloped” is based on the projections $s_x$ and  $s_y$. In addition to these projections, we introduce the following parameters,
\begin{equation}
    \Delta_{gh}=\dfrac{g^2-h^2}{g^2+h^2},
\end{equation}
and \begin{equation}
    d_{gh}=\sqrt{g^2+h^2}.
\end{equation}
The parameter $\Delta_{gh}$ measures the ellipticity, i.e. the asymmetry,  of the
intersection and $ d_{gh}$ is a distance metric for the branching space. Steeper pitched (narrower) CIs are associated with large values of $d_{gh}$. When  the upper PES and the lower PES are positioned entirely above  and below the energy of the CI, respectively, the branching plane vectors have
zero projection onto the seam coordinate, i.e., $s_x=s_y=0$. We say that the CI is peaked as it is described by vertical cones. This topographical structure closely resembles the classical image of a funnel, indicating that the population in the upper state is efficiently directed toward and through the intersection positioned at the apex of the cone \cite{Yarkony,Lee_2007}. However, the intersection with a nonzero projection on  the seam coordinate is called tilted or sloped CIs. In this scenario, the axis of the cone can be tilted to a significant degree, causing sections of the upper surface to fall below the energy of the CI, while sections of the lower surface rise above it. The symmetry and pitch of the CI can be determined by $\Delta_{gh}$. A nonzero $\Delta_{gh}$ indicates  a difference in the slopes of the PESs. Using the parameters provided in Fig.~\ref{fig_diagram}, we find that,
\begin{equation}
    s_x/d_{gh}=0.00269216,
\end{equation}
\begin{equation}
    s_y/d_{gh}=0,
\end{equation}
and 
\begin{equation}
    \Delta_{gh}=0.941036.
\end{equation}
This means that our CI is a non symmetric
sloped CI and that the cone axis is only tilted along $\textbf{x}$ axis and non tilted along $\textbf{y}$ axis. In the absence of the linear term $-{F_0{q}_x}/{2}$ in $S(Q,q)$ provided in Eq.~(\ref{S_Q_q}), we find that $s_x=s_y=0$, which means that the CI, in that case, is peaked.  In general, real CIs exhibit some degree of both tilt and asymmetry.
\section{Wavepacket and electronic dynamics}\label{Nuclear motion in  neighbourhood of a CI}
In this section, we discuss the vibronic dynamics by considering different initial states, i.e., when the system is prepared on either the lower or the upper PES.
\subsection{Initial state on the lower PES }
We first introduce the initial nuclear wavefunction, which is  a Gaussian 
wavepacket centered at $(q_x^0, Q_y^0)$, 
\begin{equation}
\label{phi}
\phi(q_x,Q_y)=\dfrac{\text{exp}\bigg[-\dfrac{1}{2}\bigg(\dfrac{q_x-q_x^0}{l^q_x}\bigg)^2-\dfrac{1}{2}\bigg(\dfrac{Q_y-Q_y^0}{l^Q_y}\bigg)^2\bigg]}{\sqrt{\pi{ l^{Q_y}}{l^{q_x}}}},
\end{equation}
where  $l^{q_x}=\sqrt{\dfrac{1}{{\mu\bar{\omega}_x}}}$ and $l^{Q_y}=\sqrt{\dfrac{1}{ M{\omega_y}}}$ are the width of the Gaussian.  In addition, we prepare the electronic state in the eigenvector $\ket{\varphi_{-}(Q,q)}$ given  in Eq.~(\ref{phi_moins}).  The total wave function of the initial state reads,
\begin{equation}
\label{phi__}
\ket{\psi_0(q_x,Q_y)}=\ket{\psi(q_x,Q_y,t=0)}=\phi(q_x,Q_y) \ket{\varphi_{-}(q_x,Q_y)}.
\end{equation} 
The time dependent state $\ket{\psi(q_x,Q_y,t)}$ is determined by solving the Schr\"odinger equation,
\begin{equation}\label{Schro}
i\dfrac{\partial}{\partial{t}}\ket{\psi(q_x,Q_y,t)}=H\ket{\psi(q_x,Q_y,t)},
\end{equation}
starting from the initial condition $\ket{\psi_0(q_x,Q_y)}$. State $\ket{\psi(q_x,Q_y,t)}$ can be expanded in terms of the diabatic  states $|\nu\rangle$ and adiabatic states $|\varphi_{\mu}(q_x,Q_y)\rangle$ as,
\begin{align}\label{population}
\notag
\ket{\psi(q_x,Q_y,t)}&=\sum_{\nu=rr,+} {\phi}_{\nu}(q_x,Q_y,t)\ket{\nu}\\&=\sum_{\mu=\pm} \tilde{\phi}_{\mu}(q_x,Q_y,t)\ket{\varphi_{\mu}(q_x,Q_y)},
\end{align}
where ${\phi}_{\nu}(q_x,Q_y,t)$ and $\tilde{\phi}_{\mu}(q_x,Q_y,t)$ are complex valued functions. The populations of the diabatic and adiabatic states can be computed  using $n_{\nu=rr,+}(t)=\int | {\phi}_{\nu}(q_x,Q_y,t)|^2 dQ_y dq_x$  and $\tilde{n}_{\mu=\pm}(t)=\int | \tilde{\phi}_{\mu}(q_x,Q_y,t)|^2 dQ_y dq_x $, respectively. To identify the effects of the CI, we shall solve the Schr\"odinger equation in the adiabatic representation and then the diabatic populations can be computed. The diabatic populations can be deduced using the relation,
\begin{equation}
\phi_{\nu}=\sum_{\mu=\pm} \tilde{\phi}_{\mu}(q_x,Q_y,t)\braket{\nu|\varphi_{\mu}(q_x,Q_y)},
\end{equation}
where $\nu=\{rr,+\}$.
{To monitor the nuclear motion, we evaluate the expectation value of the position operator $q_x$ defined by,
\begin{equation}
    \langle q_x(t)\rangle =
\iint q_x\,|\psi(q_x,Q_y,t)|^{2}\,dq_x\,dQ_y,
\label{position_operator}
\end{equation}
and its projection on the lower and upper adiabatic surfaces,
\begin{equation}
\langle q_x(t)\rangle_{\pm} =
\frac{\iint q_x\,|\tilde{\phi}_{\pm}(q_x,Q_y,t)|^{2}\,dq_x\,dQ_y}
{\iint |\tilde{\phi}_{\pm}(q_x,Q_y,t)|^{2}\,dq_x\,dQ_y}.
\label{position_operator_pm}
\end{equation}
These quantities provide a compact measure of the geometric centers of the nuclear wavepacket that evolves and distributes across the two PESs during the dynamics.
}
\begin{figure}
  \centering
\includegraphics[width=\columnwidth]{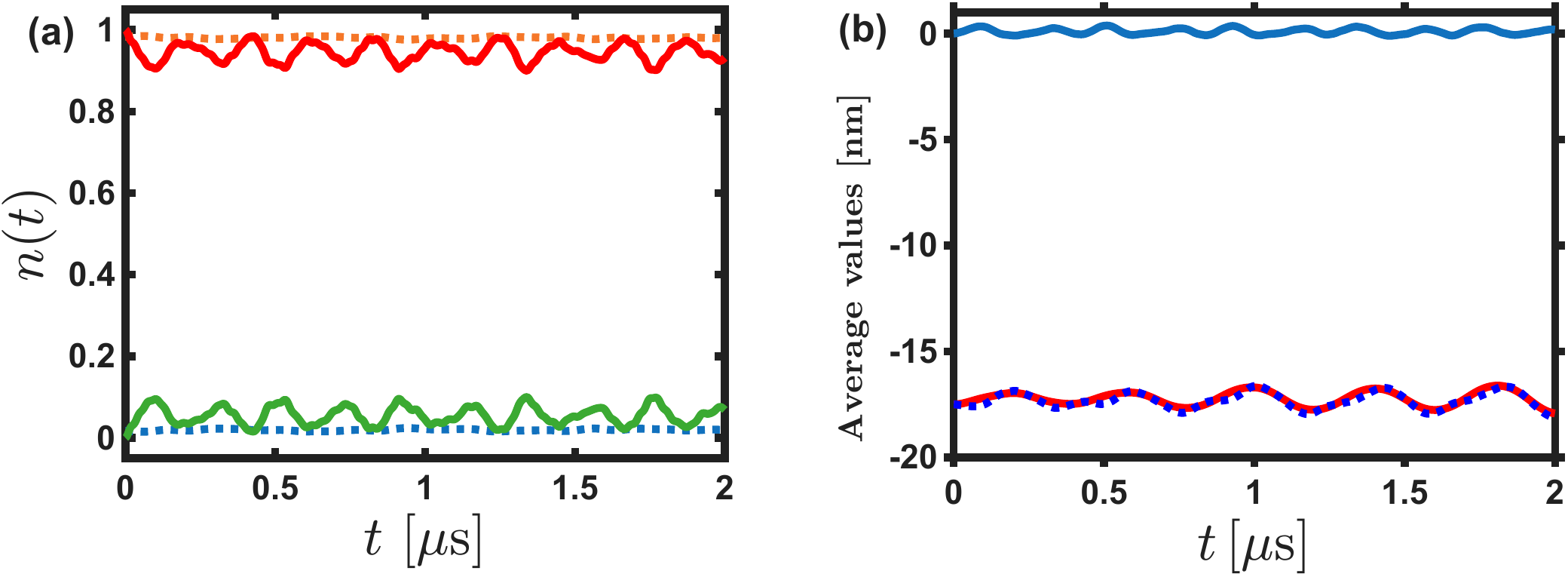}
\caption{(a) Time evolution of the diabatic and adiabatic state populations. 
The blue and orange dotted curves show the populations of the diabatic states 
$\ket{rr}$ and $\ket{+}$, respectively. 
The green and red curves show the populations of the adiabatic states 
$\ket{\varphi_{+}(Q,q)}$ and $\ket{\varphi_{-}(Q,q)}$, respectively.
(b) Time evolution of the average position $\langle q_x(t)\rangle$ (red curve), together with its 
projections on the lower PES $\langle q_x(t)\rangle_{-}$ (dotted blue curve) and upper PES $\langle q_x(t)\rangle_{+}$ (solid blue curve). Here the initial state is $\ket{\psi_0(q_x,Q_y)}=\phi(q_x,Q_y) \ket{\varphi_{-}(q_x,Q_y)}$.  $q_x^0=-17.5\,\text{nm}$, $Q_y^0=0\,\text{nm}$ and all other parameters are same as Fig.~\ref{fig_diagram}.} 
		\label{Lower_exact_-17}
\end{figure}

\begin{figure}
  \centering
\includegraphics[width=0.9\columnwidth]{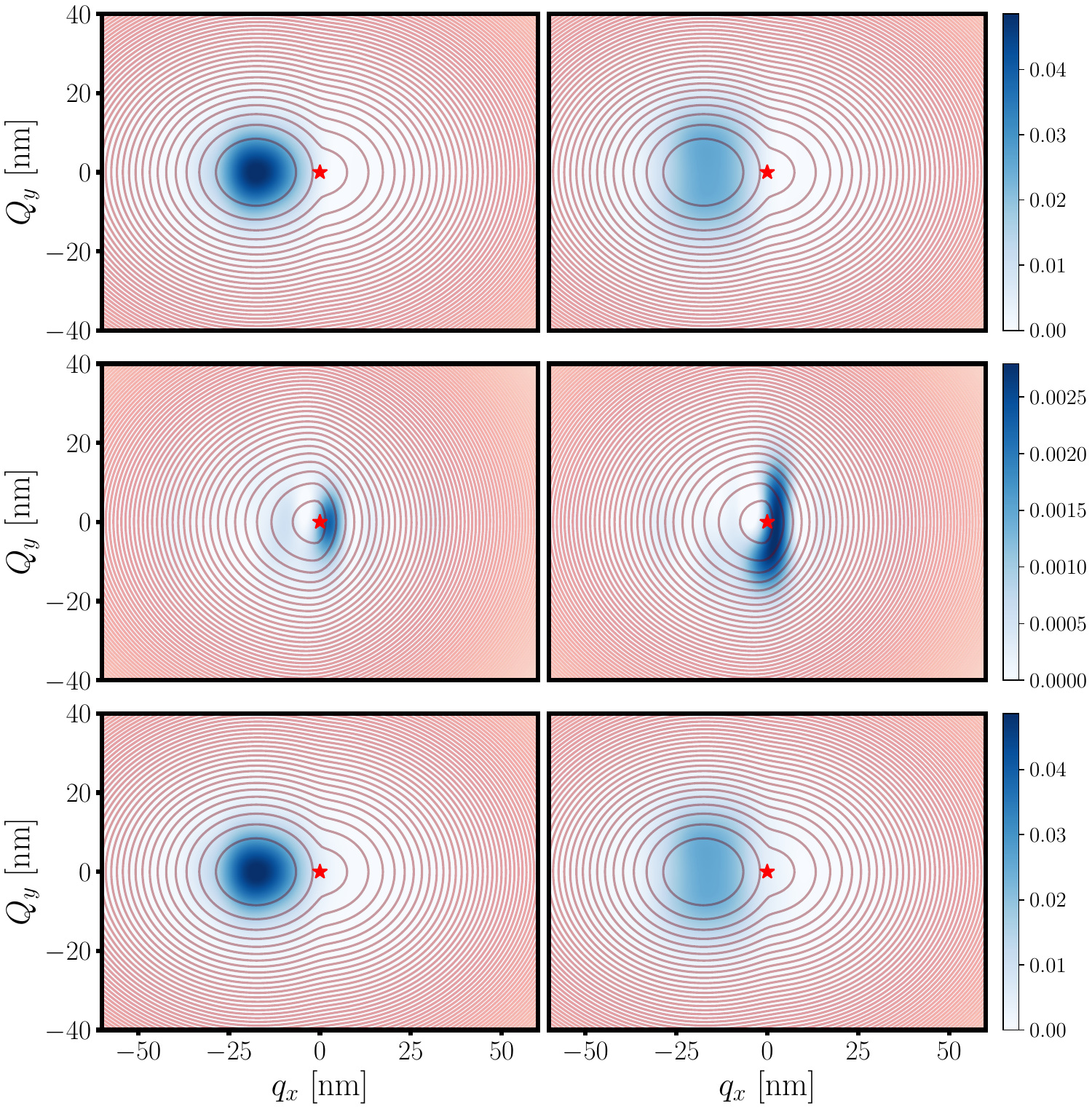}
\caption{
Dynamics of nuclear densities. 
From top to bottom: the nuclear density on the lower PES, the nuclear density 
on the upper PES, and the full nuclear density. 
Red contours indicate the shape of $E_{-}$ in the top and bottom panels, 
The middle panel indicates the shape of $E_{+}$. 
The red star marks the position of the CI. 
The left and right column correspond to snapshots at 
$t = 0.6~\mu\text{s}$ and $t = 1~\mu\text{s}$, respectively.
}		\label{Lower_WD_exact_-17}
\end{figure}
\begin{figure}
  \centering
\includegraphics[width=0.9\columnwidth]{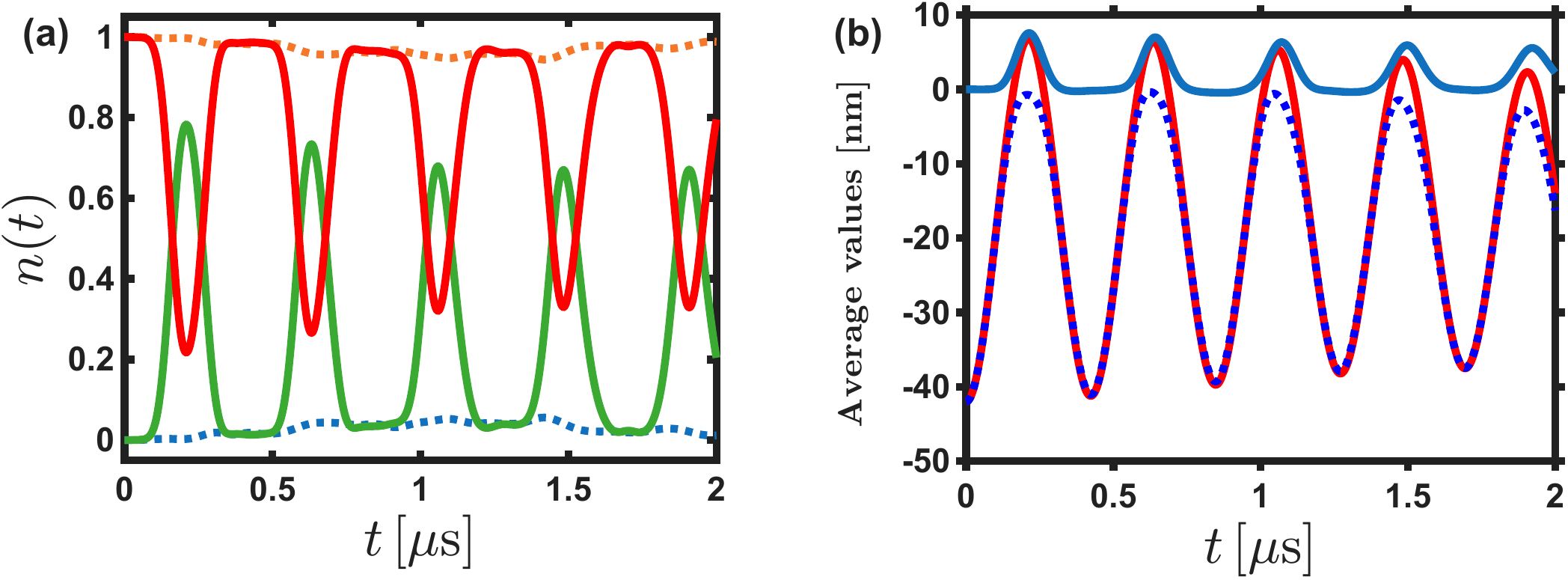}
\caption{(a) Time evolution of the diabatic and adiabatic state populations. 
The blue and orange dotted curves show the populations of the diabatic states 
$\ket{rr}$ and $\ket{+}$, respectively. 
The green and red curves show the populations of the adiabatic states 
$\ket{\varphi_{+}(Q,q)}$ and $\ket{\varphi_{-}(Q,q)}$, respectively.(b) Time evolution of the average position $\langle q_x(t)\rangle$ (red curve), together with its 
projections on the lower PES  $\langle q_x(t)\rangle_{-}$ (dotted blue curve) and the upper PES $\langle q_x(t)\rangle_{+}$ (solid blue curve). Here we have taken $\ket{\psi_0(q_x,Q_y)}=\phi(q_x,Q_y) \ket{\varphi_{-}(q_x,Q_y)}$ where $q_x^0=-42 \,\text{nm}$, $Q_y^0=0\,\text{nm}$ and all other parameters are as in Fig.~\ref{fig_diagram}.} 
		\label{Lower_exact_-42}
\end{figure}

\begin{figure}
  \centering
\includegraphics[width=0.9\columnwidth]{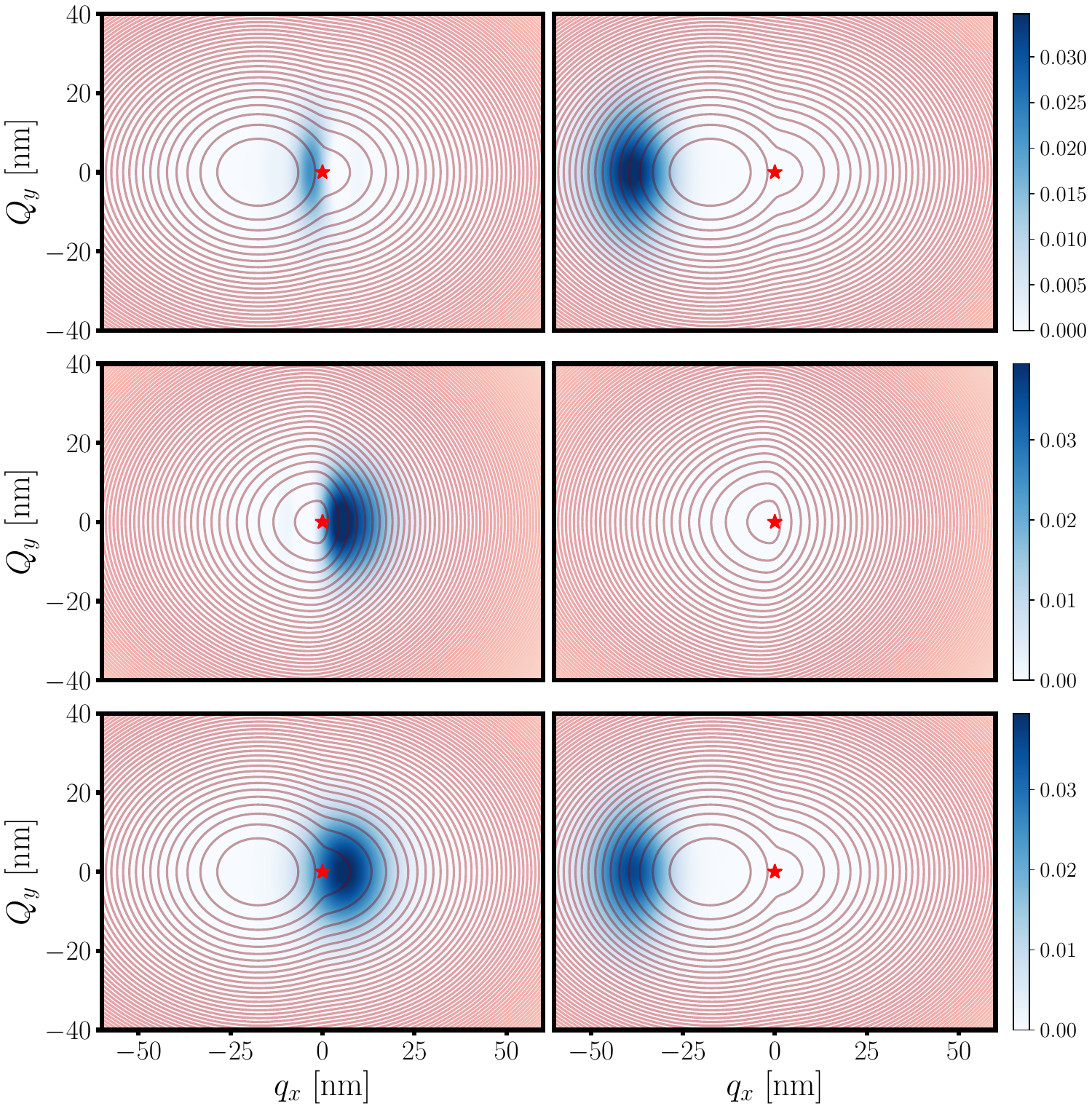}
\caption{Dynamics of the nuclear densities on the lower PES, upper PES, and the full nuclear density (top to bottom). Initially the system is prepared on the lower PES. The red contours indicate the shape of $E_{-}$ in the top and bottom panels, while in the middle panel, they indicate the shape of $E_{+}$. The red star marks the position of the CI.  The left and right columns correspond to snapshots at $t = 0.2~\mu\text{s}$ and $t = 0.4~\mu\text{s}$, respectively. Initially $q_x^0=-42 \,\text{nm}$, $Q_y^0=0\,\text{nm}$ and all other parameters are as in Fig.~\ref{fig_diagram}.} 
\label{Lower_WD_exact_-42}
\end{figure}
We first consider a case where  $q_x^0$ and $Q_y^0$ coincide with one of the two minima of the lower PES, $E_-$ i.e., $q_x^0=-17.5  \,\text{nm}, Q_y^0=0 \,\text{nm}$. Fig.~\ref{Lower_exact_-17}(a) shows the time evolution of the populations of the diabatic and adiabatic states. We find that the diabatic populations are almost  constant around their initial values, mainly for two reasons. On the one hand, the PES is not symmetric along the $q_x$-axis, and the initial state is in the potential well on the lower PES. It would cost extra energy to tunnel to the other side of the potential. During the tunneling process, on the other hand,  the wavepacket encircles the CI evenly, generating a $\pi$ phase shift. This geometric phase leads to destructive interference, such that the tunneling of the wavepacket is strongly suppressed. We find that there are few oscillations of the adiabatic populations. In particular, the population of the $\ket{\varphi_{-}(Q,q)}$ state oscillates between $0.92$ and $1$ and the population of the $\ket{\varphi_{+}(Q,q)}$ state oscillates between $0$ and $0.08$. The oscillation results from the fact that the initial state is not an eigenstate of the adiabatic basis. The data shows that the probability of finding the nuclei in the lower PES, where they are initially prepared, is higher. {To further quantify the spatial dynamics of the wavepacket, Fig.~\ref{Lower_exact_-17}(b) shows the time evolution of the average position $\langle q_x(t)\rangle$, defined in Eq.~(\ref{position_operator}), together with its components on the lower and upper PESs, $\langle q_x(t)\rangle_{-}$ and $\langle q_x(t)\rangle_{+}$, respectively.
 The quantity $\langle q_x(t)\rangle_{-}$ exhibits small oscillations around the initial value of $-17.5\,\text{nm}$, indicating that the nuclear wavepacket remains localized near the minimum of the lower PES. In contrast, $\langle q_x(t)\rangle_{+}$ changes slightly, ranging between $-0.05\,\text{nm}$ and $0.399\,\text{nm}$, reflecting the weak and transient population of the upper PES. This behavior confirms that the wavepacket stays predominantly on the lower surface, with only a minor component undergoing nonadiabatic transfer to the upper PES.
}

We now show the time slices of the wavepacket dynamics in Fig.~\ref{Lower_WD_exact_-17}. The nuclear density on the lower PES changes only slightly over time, indicating that the nuclei remain localized near their initial position and do not leave the lower PES, as shown in the top (lower PES) and bottom (total) panels in Fig. ~\ref{Lower_WD_exact_-17}. The freezing of the wavefunction in the exact dynamics is due to the intervention of the two PESs in the evolution of the system, i.e., the presence of the CI. The oscillations of $\tilde{n}_{\pm}$ are related to the excitation of the upper PES (the middle panel), where the wavepacket is weakly populated in the upper PES of the adiabatic basis during the evolution. 

Let us now consider a case in which the initial wavefunction is far from the minimum of the lower PES. Specifically, we place the wavepacket centered around $q_x^0=-42  \,\text{nm}, Q_y^0=0 \,\text{nm}$. Fig.~\ref{Lower_exact_-42} shows the time evolution of the diabatic and adiabatic populations, together with the full evolution of the position coordinate, $\langle q_x(t)\rangle$ and its average values on the upper and lower PESs $\langle q_x(t)\rangle_{\pm}$. We observe pronounced oscillations of the adiabatic populations, indicating that the system undergoes transitions between the upper and lower PESs. This behavior is also reflected in the expectation values of the position operator for the different PESs. In contrast to the case where the system is initially prepared at the minimum of the lower PES (i.e., $q_x^0 = -17.5\,\text{nm}$), the expectation values on both PESs now attain significant amplitudes due to the transition of the wavepacket between the two PESs.

In this case, the initial state has a significantly large potential energy, such that the wavepacket can tunnel between the two PESs. Fig.~\ref{Lower_WD_exact_-42} displays the corresponding nuclear densities at the two selected times. The upper, middle, and lower panels show the nuclear densities on the lower PES, upper PES, and full nuclear density, respectively. At $t = 0.2\,\mu\text{s}$, the nuclear density is clearly displaced toward the upper PES, with the distribution concentrated largely on the right-hand side of the CI (middle and bottom panels). In contrast, at $t = 0.4\,\mu\text{s}$ the density has mostly returned to the lower PES, as can be seen in both the upper PES distribution (top panel) and the total density distribution (bottom panel). This demonstrates that, unlike the situation where the system starts at the minimum of the lower PES, the nuclei are not frozen on a single energy surface but instead move and oscillate between both PESs.
\begin{figure}
  \centering
\includegraphics[width=0.9\columnwidth]{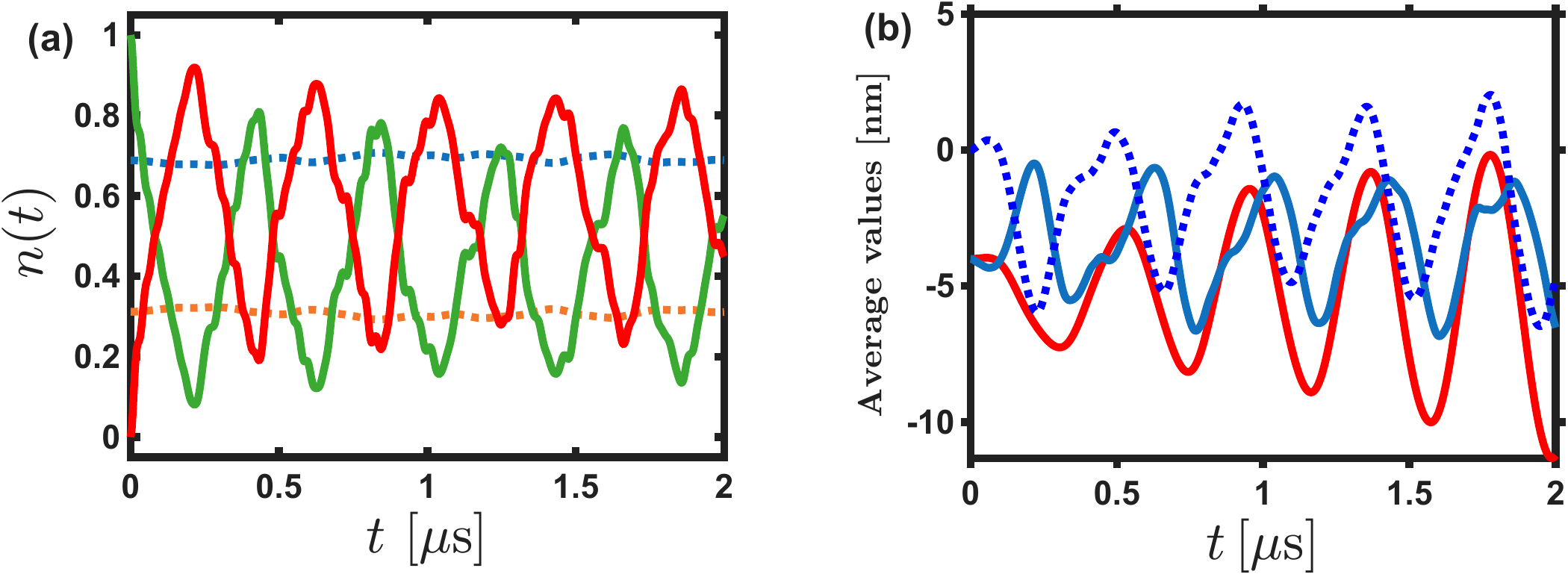}
\caption{(a) Time evolution of the diabatic and adiabatic state populations. 
The blue and orange dotted curves show the populations of the diabatic states 
$\ket{rr}$ and $\ket{+}$, respectively. 
The green and red curves show the populations of the adiabatic states 
$\ket{\varphi_{+}(Q,q)}$ and $\ket{\varphi_{-}(Q,q)}$, respectively.(b) Time evolution of the average position $\langle q_x(t)\rangle$ (red curve), together with its 
projections on the lower PES $\langle q_x(t)\rangle_{-}$ (dotted blue curve) and on the upper PES $\langle q_x(t)\rangle_{+}$ (solid blue curve). Here we have taken $\ket{\psi_0(q_x,Q_y)}=\phi(q_x,Q_y) \ket{\varphi_{+}(q_x,Q_y)}$ where $q_x^0=-4 \,\text{nm}$, $Q_y^0=0\,\text{nm}$ and all other parameters are as in Fig.~\ref{fig_diagram}.} 
		\label{Upper_exact_-4}
\end{figure}
\subsection{Initial state on the upper PES}
\begin{figure}
  \centering
\includegraphics[width=0.9\columnwidth]{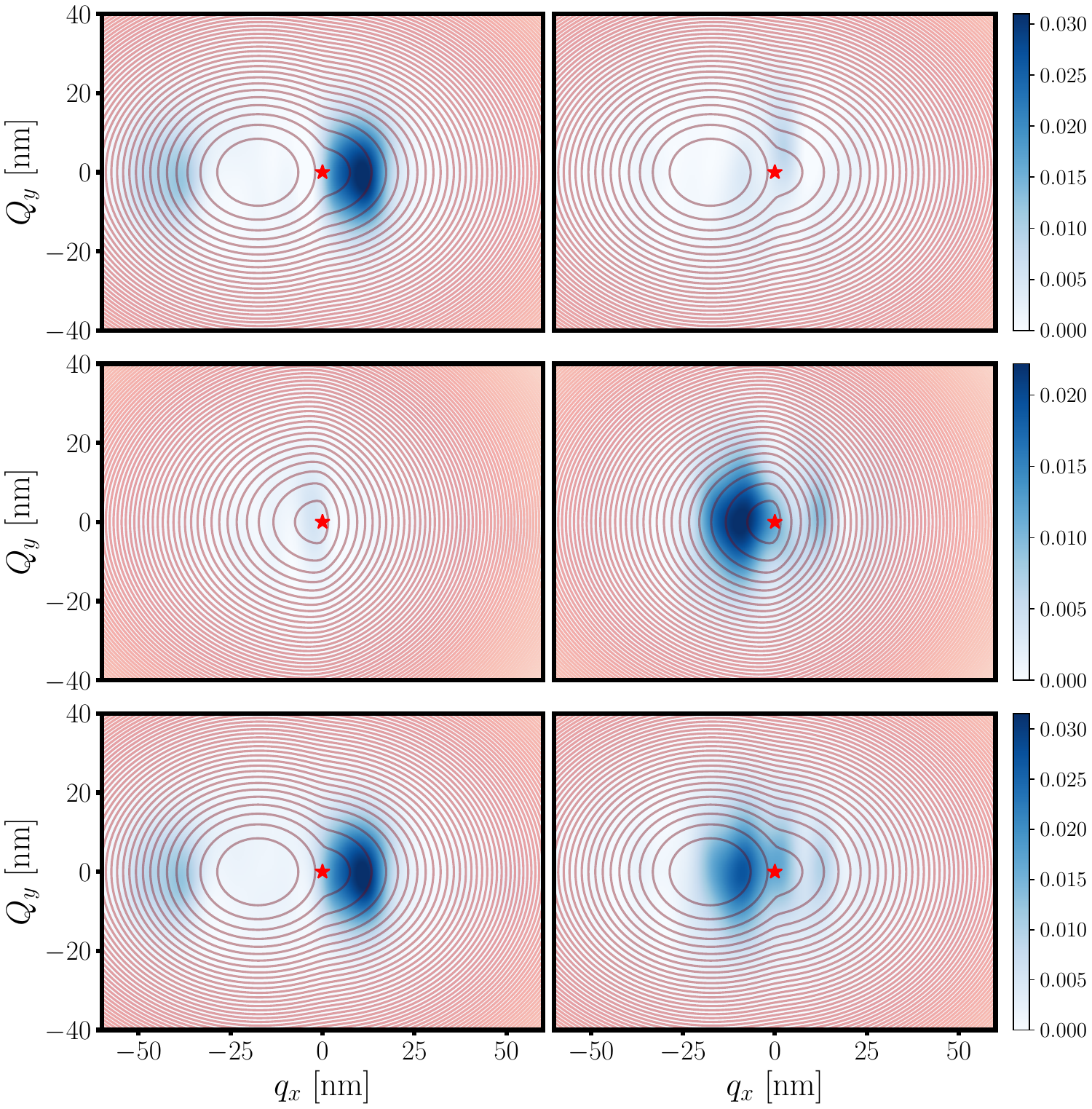}
\caption{Nuclear densities on the lower PES, upper PES, and the full nuclear density (top to bottom). Initially, the system is prepared on the upper PES of the adiabatic state. The red contours indicate the shape of $E_{-}$ in the top and bottom panels,  while in the middle panel, they indicate the shape of $E_{+}$. The red star marks the position of the CI.  Here we have taken $\ket{\psi_0(q_x,Q_y)}=\phi(q_x,Q_y) \ket{\varphi_{+}(q_x,Q_y)}$ where $q_x^0=-4\,\text{nm}$, $Q_y^0=0\,\text{nm}$ and all other parameters are as in Fig.~\ref{fig_diagram}. From left to right, the snapshots correspond to $t = 0.2~\mu\text{s}$,  and $t = 0.4~\mu\text{s}$.} 
		\label{Upper_WD_exact_-4}
\end{figure}

In this section, we investigate the dynamics of the two Rydberg ions 
in the vicinity of the CI when the system is initially prepared on 
the upper PES.
Therefore, the initial state is given by,
\begin{equation}
\label{phi_+}
\ket{\psi(q_x,Q_y,t=0)}=\phi(q_x,Q_y) \ket{\varphi_{+}(q_x,Q_y)},
\end{equation}
where $\phi(q_x,Q_y)$ is provided in Eq.~(\ref{phi}) and the electronic state is prepared in the adiabatic state  $\ket{\varphi_{+}(Q,q)}$ provided in Eq.~(\ref{phi_plus}). Let us  consider that the Gaussian state is centered near the minimum of the upper PES, i.e. $q_x^0=-4\,\text{nm}$. In Fig.~\ref{Upper_exact_-4}, we show the time evolution of the diabatic and 
adiabatic populations, together with the full evolution of the position 
coordinate $\langle q_x(t)\rangle$ and its average values on the upper and lower PESs $\langle q_x(t)_{\pm}\rangle$. 
As shown in Fig.~\ref{Upper_exact_-4}(a), the population in the diabatic basis remains almost constant in the dynamical evolution. In the adiabatic basis, on the other hand, the electronic state populations exchange between the two eigenstates. The figure also shows that $\langle q_x(t)\rangle$ oscillates around the initial value. The corresponding amplitude increases gradually as time passes, as the wavepacket oscillates. The projected $\langle q_x(t)\rangle_{\pm}$ also oscillates, where $\langle q_x(t)\rangle_{-}$ shows a phase delay with respect to $\langle q_x(t)\rangle_{+}$, which is mostly due to the initial condition. The nuclear densities, 
shown separately in Fig.~\ref{Upper_WD_exact_-4}, display the oscillatory 
character, confirming that the nuclei repeatedly migrate between the two 
surfaces during dynamics. The occupations of the upper and lower PESs are largely out of phase. When the wavepacket is largely on the lower PES at $t=0.2\mu$s (upper panel), the population on the upper PES nearly vanishes. At $t=0.4\mu$s, the density distributions of the two PESs swap.

\begin{figure}
  \centering
\includegraphics[width=0.9\columnwidth]{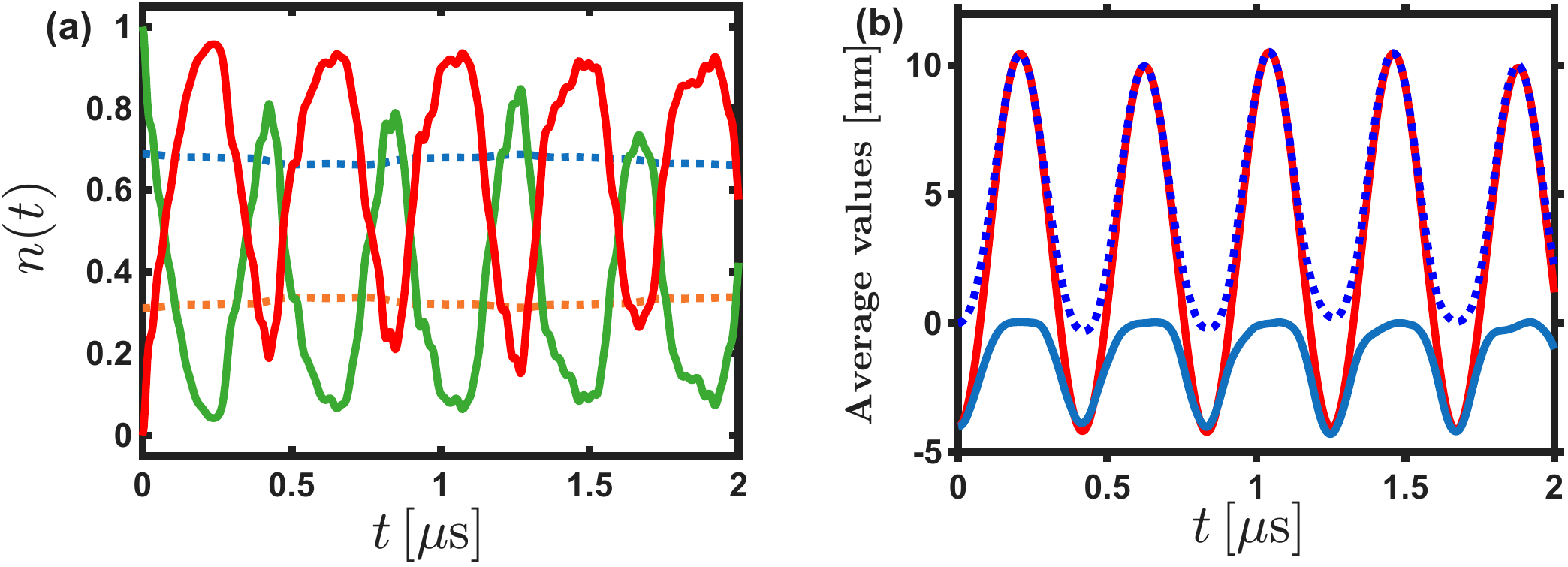}
\caption{(a) Time evolution of the diabatic and adiabatic state populations around the peaked CI. 
The blue and orange dotted curves show the populations of the diabatic states 
$\ket{rr}$ and $\ket{+}$, respectively. 
The green and red curves show the populations of the adiabatic states 
$\ket{\varphi_{+}(Q,q)}$ and $\ket{\varphi_{-}(Q,q)}$, respectively. (b) Time evolution of the average position $\langle q_x(t)\rangle$ (red curve), together with its 
projections on the lower PES $\langle q_x(t)\rangle_{-}$ (dotted blue curve) and on the upper PES $\langle q_x(t)\rangle_{+}$ (solid blue curve). Here we have taken $\ket{\psi_0(q_x,Q_y)}=\phi(q_x,Q_y) \ket{\varphi_{+}(q_x,Q_y)}$ where $q_x^0=-4 \,\text{nm}$, $Q_y^0=0\,\text{nm}$ and all other parameters are as in Fig.~\ref{fig_diagram}. Here we ignore the linear term $-F_{0}{q}_{x}/2$ in Eq.~(\ref{S_Q_q}), i.e. The CI is peaked and symmetric.} 
		\label{Upper_exact_Non_C1_-4}
\end{figure}

{Similar to the previous section, numerical simulations show that the oscillation amplitude increases when the initial wavepacket is moved away from the CI. This leads to almost complete population transfer between the two PESs coherently.}  It is worth comparing the dynamics with that of the symmetric CI as investigated in Ref.~\cite{Gambetta}. In the current system,
the asymmetry originates from the linear term $-F_{0} q_{x}/2$ provided in 
Eq.~(\ref{S_Q_q}); by neglecting this term, one recovers the peaked symmetric 
CI. To compare the properties of the two 
configurations, Fig.~\ref{Upper_exact_Non_C1_-4} 
shows the time evolution of the relevant dynamical observable for the symmetric 
(peaked) CI when the system is initially prepared on the upper PES with 
$q_x^0 = -4\,\text{nm}$. The 
oscillations of the populations on the upper and lower PESs are qualitatively 
similar in both cases. {However, the oscillations of the average position $\langle q_x(t) \rangle$ differ markedly between the two cases. For the peaked CI, the oscillations are regular and exhibit an almost constant amplitude. In contrast, the oscillations associated with the sloped CI are irregular and show a varying amplitude. For example, in the peaked CI configuration, 
$\langle q_x(t) \rangle$ oscillates over a range from 
$-4.2\,\mathrm{nm}$ to $10.5\,\mathrm{nm}$, 
while in the sloped CI configuration the oscillations are confined between 
$-11.3\,\mathrm{nm}$ and $-0.17\,\mathrm{nm}$. 
This reflects the different vibrational responses associated with these two CI geometries.

}
\section{Conclusions}\label{Conclusions}
In this paper we have proposed a new approach to engineer a sloped CI using two trapped Rydberg ions. The main particularity of our scheme is that we have  exploited the interaction between the Rydberg ions  and the sideband transition of the collective oscillation modes of the ion crystal. Compared to our previous approach~\cite{Gambetta}, an electric field is not required to push the ions away from the trap center. This could have advantages in the experimental realization, i.e. without need to changing the electric potential of the Paul trap, and avoiding electric field induced effects (such as Stark effects and ionization of the Rydberg states). We have simulated the dynamics of the nuclear wavepackets and electronic states around the CI. It is found that the sloped CI affects both the adiabatic and diabatic dynamics. Because the potential minimum is shifted from the center of the system, the geometric effect is stronger than that in the peaked CI case, where the population can be almost completely frozen in the initial state when preparing the initial wavefunction to be a Gaussian centered around the minimum of the lower PES. When initially preparing the system away from the minimum of the PES, this  results in  large oscillation amplitudes of the adiabatic populations between the two PESs. Furthermore, the effect of the intervention of the two PESs is studied by preparing the initial wavefunction on the upper PES. In this case, we find a clear swapping of the nuclear densities between the two PESs during the dynamics. In the future, it would be interesting to investigate how the CI interacts with the dissipation in a controlled manner \cite{PhysRevA.106.023309,PhysRevA.108.L050201,PhysRevX.13.011008,chaudharySpinPhononRelaxation2024}.\\\\

\textbf{Data availability:}
The data that support the findings of this article are openly available \cite{belfakir_li_2025}.
\acknowledgments 
We acknowledge support from the EPSRC through Grant No.~EP/W015641/1, and the Going Global Partnerships Programme of the British Council (Contract No.~IND/CONT/G/22-23/26). A.~B. acknowledges the hospitality of the University of Nottingham.
\bibliographystyle{apsrev4-2}
\bibliography{references}
\end{document}